\documentclass[12pt]{article}
\usepackage{epsfig}

\newcommand{\eq}{\begin{equation}}
\newcommand{\eqx}{\end{equation}}
\newcommand{\eqn}{\begin{eqnarray}}
\newcommand{\eqnx}{\end{eqnarray}}
\newcommand{\f}[2]{\frac{#1}{#2}}

\newcommand{\tr}{\mbox{\rm tr}\,}

\newcommand{\Lm}{\Lambda}
\newcommand{\lm}{\lambda}
\newcommand{\OM}{\Omega}

\newcommand{\ff}{{\cal F}}
\newcommand{\nn}{{\cal N}}
\newcommand{\qqqq}{\quad\quad\quad\quad}
\newcommand{\der}[2]{\f{\partial{#1}}{\partial{#2}}}
\newcommand{\Qt}{\tilde{Q}}
\newcommand{\cor}[1]{\left\langle {#1} \right\rangle}
\newcommand{\gam}[1]{\Gamma\!\left({#1}\right)}
\newcommand{\hyp}[6]{{}_3F_2\left({#1},{#2},{#3};{#4}, {#5}\,\Biggr|\,%
{#6}\right)}

\newcommand{\upnc}{{\cal U}^{pure}}
\newcommand{\upm}{{\cal U}^{matter}}
\newcommand{\bin}[2]{\left(\!\!%
\begin{tabular}{c}
{$#1$}\\
{$#2$}
\end{tabular}
\!\!\right)
}

\newcommand{\uf}{u^{fact.}}

\title{Explicit factorization of Seiberg-Witten curves with matter
from random matrix models.\footnote{Work partially
supported by the European Commission RTN programme HPRN-CT-2000-00131.
}\\}

\author{Yves Demasure$^{a,b}$\footnote{e-mail: {\tt demasure@nordita.dk}}\
\ and Romuald A. Janik$^{c,d}$\footnote{
e-mail: {\tt janik@nbi.dk}}\\ 
\\ \small
$^a$ Instituut voor Theoretische Fysica, Katholieke Universiteit Leuven,\\
\small Celestijnenlaan 200D, B-3001 Leuven, Belgium\\
\small $^b$ NORDITA,\\
\small Blegdamsvej 17, DK-2100 Copenhagen, Denmark\\
\small $^c$ The Niels Bohr Institute,\\
\small Blegdamsvej 17, DK-2100 Copenhagen, Denmark\\
\small $^d$ Jagellonian University,\\
\small Reymonta 4, 30-059 Krakow, Poland}

\begin{document}
\begin{titlepage}

\rightline{KUL-TF-2002-18}
\rightline{NORDITA-HE-2003-26}
\setcounter{page}{0}

\vskip 1.6cm

\begin{center}

{\LARGE Explicit factorization of Seiberg-Witten curves with matter
from random matrix models.}\footnote{Work partially
supported by the European Commission RTN programme HPRN-CT-2000-00131.}  

\vskip 1.4cm

Yves Demasure$^{a,b}$\footnote{e-mail: {\tt demasure@nordita.dk}}\
\ and Romuald A. Janik$^{c,d}$\footnote{
e-mail: {\tt janik@nbi.dk}}\\ 
\vskip 0.3cm
 
{\small
$^a$ Instituut voor Theoretische Fysica, Katholieke Universiteit Leuven,}\\
{\small Celestijnenlaan 200D, B-3001 Leuven, Belgium}\\
{\small $^b$ NORDITA,}\\
{\small Blegdamsvej 17, DK-2100 Copenhagen, Denmark}\\
{\small $^c$ The Niels Bohr Institute,}\\
{\small Blegdamsvej 17, DK-2100 Copenhagen, Denmark}\\
{\small $^d$ Jagellonian University,}\\
{\small Reymonta 4, 30-059 Krakow, Poland}

\end{center}

\vskip 1.4cm

\begin{abstract}
Within the Dijkgraaf-Vafa correspondence, we study the complete
factorization of the Seiberg-Witten curve for
$U(N_c)$ gauge theory with $N_f<N_c$ massive flavors.
We obtain explicit expressions, from random matrix theory, 
for the moduli, parametrizing the curve. 
These moduli characterize the submanifold of the Coulomb branch
where all monopoles become massless.
We find that the matrix model reveals some nontrivial structures of
the gauge theory. In particular the moduli are additive with respect
to adding extra matter and increasing the number of colors. 
\end{abstract}
\end{titlepage}

\tableofcontents

\section{Introduction}

Very recently there appeared new powerful methods of extracting
effective superpotentials for a wide class of $\nn=1$ gauge
theories. The proposal by Dijkgraaf and Vafa \cite{DV1,DV2}, building upon
earlier string theoretical constructions \cite{GEOM,CIV}, links these
superpotentials with quantities in random matrix models. The proposal
have since been proven \cite{ZANON,DCSW}. 
The proposal has been extended to theories with fundamental
matter \cite{MATTER} and there has been significant work on studying
some features of the link with Seiberg-Witten curves
\cite{CV,FERRARI,DVP,SCHNITZER1,SCHNITZER2,MOROZOV,FENG}.

These curves made their appearance in the ground-breaking work of Seiberg and
Witten \cite{SW1,SW2} on the study of $\nn=2$ supersymmetric gauge
theories. It turned out that one can describe
the low energy dynamics of the gauge theory in terms
of geometrical properties of the Seiberg-Witten
curves. Subsequently it was realized \cite{SW2,DS} that one could also
study, within
the same framework, deformations to $\nn=1$ theories by adding a
tree level superpotential. In this context one is led to
the points (submanifolds) in moduli space
where monopoles become massless and condense. At these points the
Seiberg-Witten curve
factorizes -- it has only two single zeroes (branch points) in the
case of complete factorization, where all monopoles condense. 
Once the form of the Seiberg-Witten curve at the factorization point
is known one can calculate effective potentials for the deformed
theory \cite{SW2,DS,CIV}. 

The possibility of integrating in the glueball field $S$ was realized
in \cite{FERRARI} and used to check the matrix model result following
from the Dijkgraaf-Vafa proposal for deformed pure $\nn=2$ theories
(without fundamental matter). There the explicit factorization of
curves without matter of \cite{DS} was known. Factorization properties
were also used in the context of $SO(N_c)$ theories \cite{OBERS} and
multi-trace operators \cite{deBOER} to obtain the appropriate
effective superpotentials $W_{eff}(S)$. 

However once one adds fundamental matter to the theory, the
factorization problem becomes exceedingly complicated (even in the
case of $U(2)$ with 1 flavour) and no explicit solutions are known.
The aim of this paper is to use the Dijkgraaf-Vafa proposal linking
superpotentials to matrix models and derive, from the random matrix
model solution, explicit factorization of Seiberg-Witten curves of
$U(N_c)$ theory with $N_f<N_c$ fundamental matter fields.

If it were not for the Dijkgraaf-Vafa correspondence, we would not
expect that any analytical solutions to this problem could exist. 
The fact that they can be obtained in this way shows that the random
matrix model {\em with matter} is intimately linked to fine details of
the geometry of the appropriate Seiberg-Witten curves.

In addition we find that the matrix theory variables capture a
surprising robust structure of the factorized Seiberg-Witten curve.  
The `pure' $\nn=2$ solutions and the new contributions of each
flavor appear additively. All nonlinearity is concentrated in a single
relation involving the scale of the gauge theory $\Lm$ and the matrix
theory variables. Moreover we find that the `integrating-out'
equations of {\em pure}  $\nn=2$ theories reappear, in terms of matrix
variables, in the context of $\nn=2$ theories with fundamental flavors. 
This arises naturally in the matrix model but is rather unexpected
from a gauge theory perspective.

In section 7 we collect the final results of the paper. 

The plan of this paper is as follows. In section 2 we describe the
field theoretical ingredients in more detail. In section 3 we 
introduce the setup for the Dijkgraaf-Vafa proposal with matter
fields. We then go to use the orthogonal polynomial method to rederive
directly the solution of the factorization for pure $\nn=2$ theory,
which is an ingredient of the expressions for theories with matter.
In sections 5 and 6 we derive the linearity in couplings of
the superpotential and the expressions for factorization. 
Section 7 contains the main result of the paper. In section 8 we give
some specific examples and close the paper with a discussion. 
Three appendices contain some technical details. 

\section{Field theory considerations}

It has been known for a long time that $\nn = 2$ supersymmetric
$U(N_c)$ gauge theories with $N_f < N_c$ massive flavors has a Coulomb
branch that is not lifted by quantum corrections \cite{SW1}. 
This quantum moduli space is $N_c$-dimensional, parametrized for
example by  $u_k=\cor{\frac{1}{k}\tr \Phi^k}$ with $k \leq N_c$.
At each point of the moduli space, the low energy theory
is described by an $\nn = 2$ effective abelian $U(1)^{N_c}$  gauge
theory.  
All the relevant quantum corrections in the IR can be recast in terms
of the period matrix of a particular meromorphic one-form of the auxilliarly 
complex curve --- the Seiberg-Witten curve \cite{SW1,SW1,fa,Klemm}, or more
precisely a family of genus $N_c-1$ hyperelliptic Riemann surfaces: 
\eq
\label{e.swmat}
y^2 = P_{N_c} (x, u_k)^2 - 4 \Lambda^{2N_c - N_f} \prod_{i=1}^{N_f}(x + m_i),
\eqx
with
\eq 
P_{N_c} (x, u_k) = \cor{ \det\left(xI - \Phi\right)} = \sum_{\alpha=0}^{N_c}
s_{\alpha} x^{N_c-\alpha}. 
\eqx
The coefficients $s_{\alpha}$ are polynomials of the $u_k$'s parameterizing
the Coulomb branch:
\eqn
&&\alpha s_{\alpha} + \sum_{k=0}^{\alpha} k s_{\alpha - k} u_k = 0 \\
&&  s_0=1\, , \quad  u_0= 0
\eqnx

One can deform this $\nn=2$ theory to a $\nn=1$ gauge theory by adding
a tree level superpotential: 
\eq
\label{e.tree}
W_{tree} = \sum_{p=1}^{N_c+1}  g_p \cdot \frac{1}{p} \tr \Phi^p.
\eqx
The classical vacuum structure is given by all possible 
distributions of the $N_c$ eigenvalues
$\phi_k$ of $\Phi$ amongst the $N_c$ critical points $a_j$ of the potential. 
This corresponds, at the classical level, to a breaking
of the $U(N_c)$ gauge symmetry to the  gauge group
$\prod_{i=1}^{N_c-n} U(N_i)$ with $\sum_{i=1}^{N_c-n} N_i = N_c$,
where the $N_i$'s are the nonzero multiplicities of the eigenvalues. 
For the purposes of this paper we can safely set $g_{N_c+1}=0$ as we
will be considering the case with no breaking of gauge symmetry
(complete factorization).  

Turning to the quantum picture,
the presence of this superpotential will lift the quantum moduli space,
characteristic of the
$\nn=2$ Coulomb phase,  except for the 
codimension $n$ submanifolds, where $n$ mutually local magnetic
monopoles become massless. 
These are the $\nn=1$ vacua solving the F-flatness and D-flatness conditions 
and are characterized by a monopole condensate of the massless
monopoles. This is believed to produce, by the dual Meissner effect,
the expected confinement of the electric $\nn=1$ theory. Hence, the
final quantum theory is described at low energies by a $\nn=1$
$U(1)^{N_c-n}$ gauge theory. These $U(1)$'s can be thought of as the
$U(1) \subset U(N_i)$ of the classical theory. The $\nn=1$ $SU(N_i)$
part of the theory confines, has a mass gap and is characterized by a
gaugino condensate. 

These $\nn=1$ vacua, being the codimensional $n$ submanifolds of the
$\nn=2$ Coulomb branch, where $n$ mutually local monopoles
become massless, are parameterized by the sets of moduli $\{\uf_k\}$
where the Seiberg-Witten curve factorizes, i.e. the r.h.s. of
(\ref{e.swmat}) has $n$ double roots\footnote{For simplicity we are
assuming that higher order roots are not occuring.} and $2(N_c-n)$ single
roots: 
\eq
P_{N_c} (x, \uf_k)^2 - 4 \Lambda^{2N_c - N_f} \prod_{i=1}^{N_f}(x + m_i) =
F_{2(N_c-n)}(x) H_n^2(x).
\eqx
Moreover it is shown in \cite{CV} that the reduced curve,
\eq
y^2 = F_{2(N_c-n)}
\eqx
captures the full quantum dynamics of the $\nn=1$ $U(1)^{N_c-n}$ low
energy theory. 

The effective superpotential of the theory deformed by (\ref{e.tree}),
where we set $g_{N_c+1}=0$,
is obtained by plugging the solutions $\uf_k$, parametrized by $N_c-n$
parameters into the tree level superpotential: 
\eq 
\label{qep}
W_{eff} = \sum_{p=1}^{N_c} g_p \uf_p,
\eqx
Then (\ref{qep}) should be minimized with respect to the $N_c-n$ parameters.

\subsubsection*{Complete factorization}

In this paper we will be interested in the case where $N_c-1$ mutually local 
monopoles condense. 
This corresponds to a complete factorization of the Seiberg-Witten
curve: the vacua form a 1 dimensional submanifold such that the curve
has only $2$ single roots and $N_c-1$ double roots:
\eq
P_{N_c} (x, \uf_k)^2 - 4 \Lambda^{2N_c - N_f} \prod_{i=1}^{N_f}(x + m_i) =
(x-a)(x-b) H_{N_c-1}^2(x)
\eqx

The main goal of this paper is to find explicit expressions for the
moduli $\uf_k$ where complete factorization occurs.  Let us briefly review
the solution for the {\em pure} $\nn=2$ Yang-Mills case found by
Douglas and Shenker \cite{DS} some time ago using Chebyshev polynomials.


Their solution factorizing the curve $P_{N_c}(x,u_k)^2-4\Lm^{2N_c}$
is\footnote{Adapted to the $U(N_c)$ gauge group \cite{FERRARI}.}
\eq
\label{e.ufppure}
\uf_p(\Lambda, u_1) =
\frac{N_c}{p} \sum_{q=0}^{[p/2]} \bin{p}{2q} \bin{2q}{q} \Lambda^{2q}
\left(\frac{u_1}{N_c} \right)^{p-2q}. 
\eqx
Note that the one dimensional submanifold where the $U(N_c)$ curve completely
factorized, is parametrized by $u_1 = \tr \Phi$. These results can be easily
restricted to the $SU(N_c)$ case, by putting explicitly $u_1=0$. All parameters
are then uniquely fixed in terms of the only scale of the theory $\Lambda$.

\subsubsection*{Integrating in $S$}

The quantum $\nn=1$ effective potential generated by the tree level
potential is obtained by minimizing (\ref{qep}) 
\eq 
\label{qepz}
W_{eff} (\Lambda, u_1) = \sum_{p=1}^{N_c} g_p \uf_p(\Lambda, u_1)
\eqx
with respect to $u_1$.

Here we follow an alternative route taken in \cite{FERRARI}, and 
integrate in $S$ by performing a Legendre
transformation with respect to $\log \Lambda^{2N_c-N_f}$. The superpotential
is then given by:
\eqn
\label{e.intin}
W_{eff} (S, u_1, \OM, \Lm) \!\!\!&=&\!\!\! S \log \Lambda^{2N_c-N_f}
+W_{eff}(S,u_1,\OM) = \nonumber\\
\!\!\!&=&\!\!\! S \log \Lambda^{2N_c-N_f} - S \log \OM^{2N_c-N_f} +
\sum_{p=1}^{N_c} g_p \uf_p (\OM, u_1) .
\eqnx
Note that the only $\Lambda$-dependence is in the linear $S \log
\Lambda^{2N_c-N_f}$ term.  
Integrating out $S$ forces $\OM=\Lambda$ and brings us back to
(\ref{qepz}). In order to get the effective potential
$W_{eff}(\Lambda,S)$ one has to integrate out both $\OM$ and
$u_1$.

Our strategy for identifying the factorization parameters $\uf_p$ for
the theory with matter is to rewrite the random matrix expression in
the form given by the last two terms of (\ref{e.intin}) and to
read off the appropriate gauge theoretic moduli $\uf_p$.
The first term in (\ref{e.intin}) could also be absorbed into the
matrix model expressions if appropriate rescalings of the integration
measure of the matrix model was made. We will not do this here.


For later reference let us quote the equations of motion for the pure
$\nn=2$ gauge theory: 
\eqn
\label{e.weffom}
\frac{\partial W_{eff}(S,u_1, \OM)}{\partial  \log \OM^{2N_c}}
\!\!\!&=&\!\!\!  
\sum_{p \geq 2} g_p   \sum_{q=0}^{[p/2]} \frac{q}{p}
\bin{p}{2q} \bin{2q}{q} \OM^{2q} \left(\frac{u_1}{N_c} \right)^{p-2q} -
S = 0 \\
\label{e.weffu}
\frac{\partial W_{eff}(S,u_1, \OM)}{\partial u_1} \!\!\!&=&\!\!\! 
\sum_{p \geq 1} g_p   \sum_{q=0}^{[p/2]} \frac{p-2q}{p}
\bin{p}{2q} \bin{2q}{q} \OM^{2q} \left(\frac{u_1}{N_c}
\right)^{p-2q-1} \!\!\! = 0
\eqnx
These follow directly from (\ref{e.intin}) and (\ref{e.ufppure}).


\section{The Dijkgraaf-Vafa proposal with fundamental matter}

According to the Dijkgraaf-Vafa prescription the the perturbative part of the
superpotential can be expressed as \cite{DV1}
\eq
\label{e.weffmat}
W_{eff}(S)=N_c \der{\ff_{\chi=2}(S)}{S} +\ff_{\chi=1}(S)
\eqx
where the $\ff_\chi$ are defined through the matrix integral
\cite{MATTER,DV1} 
\eq
\label{e.rmmat}
e^{-\f{N^2}{S^2} \ff_{\chi=2}(S)-\f{N}{S} \ff_{\chi=1}(S)+\ldots} =
\int D\Phi DQ_i D\Qt_i e^{-\f{N}{S} \left( \tr V(\Phi) -m Q_i \Qt_i
-Q_i\Phi \Qt_i \right)}
\eqx
where we included the coupling of fundamental matter to the adjoint
field in accordance with $\nn=2$ supersymmetry. 

The matter fields in the matrix model (\ref{e.rmmat}) appear only
quadratically and hence may be integrated out giving 
\eq
\label{e.quadr}
\int D\Phi \det \left(m+\Phi\right)^{-1} e^{-\f{N}{S} \tr V(\Phi)}=
Z \cdot \cor{\det \left(m+\Phi\right)^{-1}}
\eqx
where $Z$ is the partition function of the matrix model without
matter. For the complex matrix model it is well know that $Z$ will
contribute only to $\ff_{\chi=2}$ and not to $\ff_{\chi=1}$.
Using large $N$ factorization we also have
\eq
\label{e.largenf}
\cor{\det (m+\Phi)^{-1}}=\cor{e^{-\log \det (m+\Phi)}}
=\cor{e^{-\tr \log(m+\Phi)}}= e^{-\cor{ \tr \log(m+\Phi)}} 
\eqx
We see that the matter determinant appears at subleading order in $N$
w.r.t the tree level potential. This has the important consequence
that the saddle point solution (eigenvalue density $\rho(\lm)$) and
hence $\ff_{\chi=2}$ will {\em not be influenced} by the presence of
matter. The $\chi=1$ contribution is then given by
\eq
\label{e.rmchi}
\ff_{\chi=1}= \sum_{i=1}^{N_f} \int d\lm \rho(\lm) \log(m_i+\lm)
\eqx

In the following section we will evaluate the $\ff_{\chi=2}$ piece of
the partition function $Z$ appearing in (\ref{e.quadr}) using the
method of orthogonal polynomials. As a byproduct we will rederive the
factorization expression for the pure gauge theory case. Then in
section 5 we will start investigating the matter contribution
(\ref{e.rmchi}). 

\section{Orthogonal polynomials and $\ff_{\chi=2}$}

In this section we will use the method of the orthogonal polynomials
to study thoroughly the one cut solution. A very nice introduction 
to this powerful method can be found in \cite{fgz}.
We will show that all results obtained
from the field theory analysis appear naturally in this setting from
random matrix computations (compare \cite{FERRARI}). 

The orthogonal polynomials associated to the matrix model with
potential $\f{N}{S} \tr V(\Phi)$ satisfy the recursion
relation
\eq
s P_n(s)=P_{n+1}(s) + T_n P_n(s) +R_n P_{n-1}(s)
\eqx
In the large $N$ limit the recursion coefficients $T_n$, $R_n$ can be
taken to be continous functions of $u=S\cdot (n/N)\equiv S \cdot x$.  
In addition, the equations of motion of the matrix model
with general potential $W_{tree}(\Phi)$ become purely algebraic
\cite{GrossMigdal,fgz}:  
\eqn
\int \frac{dz}{2 \pi i} V' \left( z + \frac{R(Sx)}{z} +
T(Sx) \right) &=& S x \\
\int \frac{dz}{2 \pi i} \f{1}{z} V' \left( z + \frac{R(Sx)}{z} + T(Sx)
\right) &=& 0  
\eqnx
When $x=1$ we will denote $R(S)$ by $R$ and $T(S)$ by $T$.
These two variables are  
related to the endpoints $a$ and $b$ of the support of the matrix
eigenvalue distribution through 
\eqn 
T &=& \frac{a+b}{2} \\ 
R &=& \frac{(a-b)^2}{16}.
\eqnx
For a general matrix potential $V(\Phi)=\sum g_p \f{1}{p}\Phi^p$, one
obtains easily: 
\eqn 
\label{eomu}
u &\equiv& Sx =  \sum_{p \geq 2} g_p   \sum_{q=0}^{[p/2]}
\frac{q}{p} \bin{p}{2q} \bin{2q}{q} R(Sx)^{q} T(Sx)^{p-2q} \\
\label{eomv}
v &\equiv& \sum_{p\geq 1} g_p   \sum_{q=0}^{[p/2]} \frac{p-2q}{p}
\bin{p}{2q} \bin{2q}{q} R(Sx)^{q} T(Sx)^{p-2q-1} = 0, 
\eqnx
defining $u$ and $v$ for later convenience\footnote{Apart from the
present section $u$ and $v$ are always taken at $x=1$.}.
In terms of these polynomials, the matrix model partition function takes
a very elegant form:
\eq 
\label{Zn}
Z = N ! h_0^N  e^{N^2 \int_{0}^{1} dx (1 - x) \ln R(Sx)}.
\eqx
where $h_0$ is the integral
\eq
h_0=\int_{-\infty}^\infty ds \, e^{-\f{N}{S} V(s)}
\eqx
{}From equation (\ref{Zn}), one can easily extract the $\chi = 2$
contribution (we will comment on the $h_0$ piece below)
\eq
\ff_{\chi=2} = - S^2 \int_{0}^{1} dx (1 - x) \ln R(Sx)
\eqx
Performing an integration by parts leads to:
\eq
- S \int_{0}^{S} du \ln R(u) + \int_{0}^{S} du u \ln R(u)
\eqx
Following the Dijkgraaf-Vafa prescription, the contribution to
the gauge theory effective potential is proportional to:
\eq 
\label{int}
\frac{\partial \ff_{\chi=2}}{\partial S} = - S \ln R(S) + 
\int_{R(0)=0}^{R(S)} \frac{ u(R,T)}{R} dR
\eqx
The remaining integral is rather tricky. The integrand is a function of
two variables $R$ and $T$, tied together by the highly nonlinear
constraint $v=0$. 
This complicates the integral, and moreover spoils the manifest linearity in
the couplings $g_p$, appearing in the matrix potential.
It is convenient, however, to treat the one-dimensional integral as an
integral of a 1-form over the path $\{v=0\}$ in the two dimensional
$R$-$T$ plane.
Then one can rewrite the integral as an integral
over a {\em closed} 1-form $\omega$, such that the original integral
remains unchanged. 
One can then deform the contour $v=0$ into a  piece with $R=0$, while
integrating $T$ from $T(0)$ to $T(S)$ and another piece, keeping $T$ fixed
at $T(S)$, and integrating $R$ from $0$ to $R(S)$ (see fig. 1):
\eq
\int_{R(0)=0}^{R(S)} \frac{ u(R,T)}{R} dR \equiv \int_{v=0}
\f{u(R,T)}{R} dR= \int_{v=0} \omega =
\int_{T=const} \omega +\int_{R=const} \omega
\eqx

\begin{figure}[bt]
\centerline{%
\epsfysize=5cm 
\epsfbox{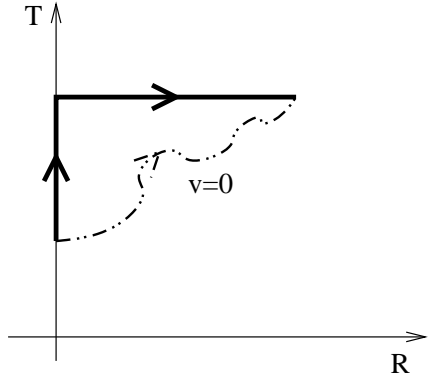}}
\caption{The dotted line represents the original contour of
integration, while the solid lines show the deformed contour in the
$T$-$R$ plane.}
\end{figure}

\noindent{}An extension of the integrand of (\ref{int}) to a closed
one-form can be found to be
\eq 
\omega=\frac{u}{R} dR + v dT.
\eqx
Note that the extra piece doesn't give a contribution to
the original integral, which is taken over $v=0$.

Performing the integral is now straightforward. The first piece, 
integrating over $T$ and constraining $R$ to $0$ gives:
\eq
\int_{T(0)}^{T(S)} v \, d T = \sum_{p \geq 1} \frac{g_p}{p} \,
\big(T^p (S) - T^p (0) \big).
\eqx
The evaluation of the integral at $T(0)$, can be cancelled (at the
saddle point) against
the $h_0^N$, appearing in the partition fuction (\ref{Zn}).
For the second piece, one fixes $T$ at $T(S)$, and integrates over $R$:
\eq
\int_{0}^{R(S)} \frac{u}{R}\, d R = \sum_{p \geq 2}  
\frac{ g_p}{p}  \sum_{q=1}^{[p/2]}
\bin{p}{2q} \bin{2q}{q} R^{q} T^{p-2q}
\eqx
{}From now on until the end of the paper we will denote by $R$ and $T$
the recursion coefficients defined at $S$. 
Bringing the two contributions together, leads to the following result:
\eq
\label{e.purefin}
W_{eff} = - N_c S \ln R + N_c \sum_{p \geq 1}  
\frac{ g_p}{p}  \sum_{q=0}^{[p/2]}
\bin{p}{2q} \bin{2q}{q} R^{q} T^{p-2q}
\eqx
This is exactly, term by term, the field theory result for {\em pure}
$\nn=2$ $U(N_c)$ gauge theory ((\ref{e.intin}) and (\ref{e.ufppure})),
provided we identify field theory variables and matrix theory variables in the
following way: 
\eq
\label{e.pureident}
R = \OM^2, \qqqq  T = \frac{u_1}{N_c} 
\eqx
The first identification is extracted from the linear term in $S$, the second
part follows from the term linear in $g_1$. As it should,
the equations of motion for $R$ and $T$ (\ref{eomu})-(\ref{eomv}), are
mapped, under this 
identification, to the equations obtained from
integrating out $\OM$ and $u_1$ (\ref{e.weffom})-(\ref{e.weffu}).

{}From (\ref{e.purefin}) we may read off the coefficients of $g_p$:
\eq
\label{e.upnc}
\upnc_p(R,T)= \f{1}{p} \sum_{q=0}^{[p/2]} \bin{p}{2q} \bin{2q}{q}
R^{q} T^{p-2q} 
\eqx
Under the identification (\ref{e.pureident}) and setting $\OM=\Lm$ 
these coincide exactly
with the factorization solution (\ref{e.ufppure}) of Douglas and
Shenker for pure $\nn=2$ gauge theory. For theories with matter,
(\ref{e.upnc}) will be just a part of the final result, and an
identification between $R$, $T$ and gauge theoretic quantities will
have to be made only after the $\ff_{\chi=1}$ contribution is evaluated in
the following sections.


\section{The $\ff_{\chi=1}$ matter contribution}

In the previous section we have rederived the result that the first
piece in (\ref{e.weffmat}) can be recast in the form
\eq
\label{e.pure}
N_c \der{\ff_{\chi=2}}{S}= N_c \left[ -S \log R +\sum g_p\,
\upnc_p(R,T) \right]  
\eqx
and the equations of motion for $R$ and $T$ derived from (\ref{e.pure})
are exactly the random matrix saddle point equations
\eqn
\label{e.eomu}
S &=& u \equiv \sum_{p \geq 2} g_p   \sum_{q=0}^{[p/2]} \frac{q}{p}
\bin{p}{2q} \bin{2q}{q} R^{q} T^{p-2q} \\
\label{e.eomv}
0 &=& v \equiv \sum_{p\geq 1} g_p   \sum_{q=0}^{[p/2]} \frac{p-2q}{p}
\bin{p}{2q} \bin{2q}{q} R^{q} T^{p-2q-1}
\eqnx
The first of these can be interpreted as the formula for
integrating in $S$. The linearity of (\ref{e.pure}) in the random
matrix couplings is essential for identifying the $\upnc_p(R,T)$ with
the point in $\nn=2$ moduli space where the Seiberg-Witten curve
$y^2=P_{N_c}(x,u_k)^2 -4\Lm^{2N_c}$ factorizes.

In the following section we will recast the random matrix expression
(\ref{e.rmchi}) for $\ff_{\chi=1}$ in the same way:
\eq
\label{e.new}
\ff_{\chi=1}=\sum_{i=1}^{N_f} \left[ S \log L(R,T,m_i) +\sum g_p\,
\upm_p(R,T,m_i) \right]
\eqx

Let us comment on an ambiguity, leading to a unique determination of a
crucial term to obtain the correct final result.
Once we start interpret this expression for $\ff_{\chi=1}$, as a part of 
the effective gauge theory potential, we should be able to integrate out 
$R$ and $T$. The equations of motion thus obtained, should be
consistent with the random matrix saddle point equations.
But, as explained in section 3, adding fundamental matter in the gauge theory
does not modify the saddle point equations of the matrix model.
So it is necessary to impose that the equations of motion for $R$ and $T$,
derived from the effective potential, with the $\ff_{\chi=1}$ 
contribution  are consistent with (\ref{e.eomu}) and (\ref{e.eomv}).
This fact is highly non-trivial and reveals an unexpected relation
between the factorization of Seiberg-Witten curves with and without matter.

We will show that the addition of $\ff_{\chi=1}$ to the superpotential
does not change the equations of motion for $R$ and $T$ provided we add
an extra term proportional to $v$.
From the matrix model perspective nothing changes, while 
the extra term turns out to be crucial to obtain the correct gauge theoretical
interpretation of our results.
We stress that this does not alter the Dijkgraaf-Vafa prescription. It
merely solves an ambiguity that arises when interpreting the matrix
model variables directly in terms of gauge theoretical quantities.




Let us now focus on a single summand of (\ref{e.rmchi}). In this case
it seems to be difficult to use the orthogonal polynomial method, which
was the most straightforward way of deriving the $\ff_{\chi=2}$
contribution. Here it is more convenient to use the saddle point
expression for the eigenvalue density of the single-cut solution:
\eq
\rho(x)=\f{1}{2\pi} M(x) \cdot \sqrt{(b-x)(x-a)} 
\eqx
where $M(x)$ is a polynomial which is expressible in terms of the
random matrix potential through
\eq
\label{e.defmz}
M(x)=\int_{C_\infty} \f{dw}{2\pi i} \f{V'(w)}{(w-x) \sqrt{(w-a)(w-b)}}
\eqx
The features of interest of the above expression are (i) it is {\em linear}
in the explicit dependence on the couplings $g_p$, (ii) the
coefficients of $g_p$ are universal functions of the endpoints $a$,
$b$ and hence of the variables $R$ and $T$, (iii) for given $g_p$, its
coefficient is a polynomial in $x$ of order $p-2$.

In order to obtain an explicit dependence on $R$ and $T$ let us
perform the change of variables
\eq
x=\f{1}{2}(a+b)+\f{b-a}{2} \psi \equiv T+2\sqrt{R}\psi
\eqx
Then the contribution to $\ff_{\chi=1}$ of a single flavour is given by
\eq
\label{e.master}
R \cdot \f{2}{\pi} \int_{-1}^1 d\psi \sqrt{1-\psi^2}\cdot M(\psi)\cdot \log
\left( m+T+2\sqrt{R}\psi \right) + v f(R,T)
\eqx
As noted before, one should take into account a possible addition of
$v$ multiplied by any function $f(R,T)$ (since $v=0$ is an
equation of motion of the random matrix model).

\section{Factorization formulas}

We will first derive the formulas involving $g_1$
and $g_2$, in particular this will allow us to fix uniquely the function
$f(R,T)$. Also this will make the general structure more transparent.
Then we will derive the results for arbitrary $g_p$.

\subsubsection*{Contribution of $g_1$, $g_2$}

To this order $M(\psi)=g_2$ and the formula (\ref{e.master}) gives
\eq
g_2 R \left[ \log(m+T) +\f{2}{\pi} \int_{-1}^1 d\psi \sqrt{1-\psi^2}
\log \left( 1+2\f{\sqrt{R}}{m+T} \psi \right) \right]
\eqx 
This can be evaluated to give 
\eq
g_2 R \left[ \log\left(\f{m+T+\sqrt{(m+T)^2-4R}}{2}\right) + (m+T)
\f{m+T-\sqrt{(m+T)^2-4R}}{4R} -\f{1}{2} \right]
\eqx
At this stage we should identify the coefficient of the logarithm with
$S$ (here this is trivial since to this order the equation of motion
for $R$ is just $S=g_2 R$, but later we will see that this property
will hold in general). Thus we are left with
\eq
\label{e.almost}
S \log\left(\f{m+T+\sqrt{(m+T)^2-4R}}{2}\right) + g_2 \left[
\f{m+T}{4} (m+T-\sqrt{(m+T)^2-4R}) -\f{R}{2} \right]
\eqx 
which indeed has the form of (\ref{e.new}). 
Now we require that the saddle point equations remain consistent with
the $\ff_{\chi=1}$ equations of motion. This detemines uniquely the
term 
\eq
\label{e.corr}
v(g_1,g_2,R,T) \cdot f(R,T) =(g_1+g_2 T) \cdot f(R,T) 
\eqx
Indeed the requirement that integrating out $T$ from the sum of
(\ref{e.almost}) and (\ref{e.corr}) gives $v=g_1+g_2 T=0$ fixes
$f(R,T)$ uniquely to be
\eq
\label{e.frt}
f(R,T)=-\f{1}{2} \left( m+T -\sqrt{(m+T)^2-4R} \right)
\eqx
This extra term does not change the equation for $R$, which
is consistent with the saddle point equations.
This function will stay unchanged in the general case. 
We may now read off the final expressions for the mass dependent
contributions to $u_1$ and $u_2$:
\eqn
\upm_1(R,T,m) &=&  -\f{1}{2} \left( m+T -\sqrt{(m+T)^2-4R} \right) \\
\upm_2(R,T,m) &=& \f{m+T}{4} (m+T-\sqrt{(m+T)^2-4R}) -\f{R}{2} + \nonumber\\
&& -\f{T}{2} \left( m+T -\sqrt{(m+T)^2-4R} \right) 
\eqnx

\subsubsection*{Arbitrary $g_p$}

We will now extend the previous considerations to the calculation of 
arbitrary $\upm_p$. The general structure will remain unchanged. The
function $f(R,T)$ multiplying $v$ will remain unmodified (as it
should). Also the coefficient of the logarithm will turn out to be
exactly $S$.

In appendix A we derive the following expression for the polynomial
$M(\psi)$:
\eq
M(\psi) = \sum_{p \geq 2} g_p \sum_{n=0}^{p-2} c_{p,n} \psi^n
\eqx
where
\eq
\label{e.defcpn}
c_{p,n}=2^n R^{\f{n}{2}} \sum_{k=0}^{\left[ \f{p-n-2}{2} \right]}
\bin{2k}{k} \bin{p-1}{2k+n+1} R^k T^{p-n-2-2k} 
\eqx
So the integral (\ref{e.master}) can be rewritten as
\eqn
\label{e.gen}
&&\sum_{p>2} g_p \sum_{n=0}^{p-2} c_{p,n} R \biggl[\log(m+T) \cdot
\f{2}{\pi} \int_{-1}^1 d\psi \sqrt{1-\psi^2} \cdot \psi^n  + \nonumber\\
&&\ \ \ \ \ \  +\f{2}{\pi} \int_{-1}^1 d\psi \sqrt{1-\psi^2} \cdot \psi^n 
\cdot \log \left( 1+2\f{\sqrt{R}}{m+T} \psi \right) \biggr]
\eqnx
We now have to distinguish two cases.

{\bf $n$ odd:} Then the first
integral vanishes and we will denote the second integral by
$f_n(z)$. As discussed in appendix B, $f_n(z)$  is essentially a polynomial
in $z$ of order $(n+1)/2$ (divided by $(z-1)^{n/2+1}$) when expressed in
terms of the variable  
\eq
\label{e.defz}
z=\f{m+T}{2R} \left(m+T+\sqrt{(m+T)^2-4R} \right)
\eqx
Appendix B contains a general formula for $f_n(z)$. 
Explicit expressions for some specific cases are shown in table 1 in
section 7.

{\bf $n$ even:} In this case the first integral is nonvanishing. Moreover
the second integral involves a logarithm with the same coefficient as
$\log(m+T)$ in the first integral. Together they combine to give
\eq
\label{e.slog}
\left\{ \sum_{p \geq 2} g_p \sum_{l=0}^{[\frac{p-2}{2}]} c_{p,2l} R \cdot
\f{2^{-2l}}{l+1} \bin{2l}{l} \right\} \cdot
\log\left(\f{m+T+\sqrt{(m+T)^2-4R}}{2}\right) 
\eqx 
It is shown in appendix C that
the coefficient in curly braces is exactly equal to $S$.
The second integral with the logarithmic part subtracted out has
again a simple polynomial structure (see appendix B for details). 

At this stage we arrive to the analogue of (\ref{e.almost}):
\eq
\label{e.almostgen}
S \log\left(\f{m+T+\sqrt{(m+T)^2-4R}}{2}\right) + \sum_{p>2} g_p
\left[ \sum_{n=0}^{p-2} c_{p,n} R f_n(z) \right]
\eqx
Again we have to add to this the correction term
\eq
\label{e.corrgen}
v \cdot f(R,T) \equiv
-\sum_{p\geq 1} g_p v_p \f{1}{2} \left( m+T -\sqrt{(m+T)^2-4R} \right)
\eqx
We checked explicitly for some cases that this term together with
(\ref{e.almostgen}) gives equations of motion consistent with the
random matrix constraints.

The sum of (\ref{e.almostgen}) and (\ref{e.corrgen}) 
is now of the expected form (\ref{e.new}), thus
defining $\upm_p(R,T,m_i)$.

\section{Final results}

Putting all results together ((\ref{e.purefin}), (\ref{e.almostgen}),
(\ref{e.corrgen})), and putting in the term $S \log
\Lambda^{2N_c-N_f}$ as required from the Dijkgraaf-Vafa prescription   
gives a prediction from the matrix model side for the quantum
effective gauge potential: 
\eqn \label{fqep}
W_{eff}(S, T, R,\Lm) =
S \log \frac{\Lambda^{2N_c-N_f}\prod_{i=1}^{N_f} \f{1}{2}
       \left(m_i+T+\sqrt{(m_i+T)^2-4R}\right)}{R^{N_c}} \nonumber \\
+ \sum_{p \geq 1} g_p \left[ 
    N_c\, \upnc_p(R,T) + \sum_{i=1}^{N_f}\, \upm_p(R,T,m_i) \right].
\eqnx
This expression should be compared with the potential
$W_{eff}(S,u_1,\OM,\Lm)$ (see (\ref{e.intin})),
obtained from the field theory analysis. The relation between the parameters
of the matrix model and the field theory 
 is highly non-linear:
\eqn
u_1 = N_c T  - \f{1}{2} \sum_{i=1}^{N_f}
       \big(m_i+T - \sqrt{(m_i+T)^2-4R}\big)\\
\label{rtuo}
\Omega^{2N_c - N_f} = \frac{R^{N_c}}{\prod_{i=1}^{N_f} \f{1}{2}
       \left(m_i+T+\sqrt{(m_i+T)^2-4R} \right)}
\eqnx
In order to obtain the final factorization formulae we
integrate out $S$ from $W_{eff}(S,T,R,\Lm)$ :
\eq
\Omega^{2N_c-N_f} = \Lambda^{2N_c-N_f},
\eqx
Combined with (\ref{rtuo}) this gives an expression
for $\Lambda$ in terms of $R$ and $T$.
The remaining part of (\ref{fqep}) should then be compared with field
theory result: 
\eq
W_{eff} (\Lambda, u_1) = \sum_{p \geq 1} g_p \uf_p(\Lambda, u_1),
\eqx
where the $\uf_p$, is the one parameter solution, factorizing
completely the Seiberg-Witten curve for $U(N_c)$ SYM with $N_f$
flavours ($N_f<N_c$): 
\eq
y^2=P_{N_c}(x,u_k)^2 -4\Lm^{2N_c-N_f} \prod_{i=1}^{N_f} (x+m_i) 
\eqx
Comparing with the results from matrix models gives an expression
for the $\uf_p$'s,
\eq
\label{e.ufpfinal}
\uf_p = N_c\, \upnc_p(R,T) +\sum_{i=1}^{N_f}\, \upm_p(R,T,m_i)
\eqx
in terms of two parameters $R$ and $T$, tied together with the constraint:
\eq
\Lambda^{2N_c - N_f} = \frac{R^{N_c}}{\prod_{i=1}^{N_f}
       \f{1}{2}\left(m_i+T+\sqrt{(m_i+T)^2-4R}\right)}
\eqx
For completeness, we recall the formulas
\eqn
\upnc_p(R,T)\!\!\! &=&\!\!\! \f{1}{p}  \sum_{q=0}^{[p/2]}
\bin{p}{2q} \bin{2q}{q} R^{q} T^{p-2q}\\
\upm_1(R,T,m)\!\!\! &=&\!\!\! -\f{1}{2} \left( m+T -\sqrt{(m+T)^2-4R}
\right) \\ 
\upm_{p\geq 2}(R,T,m)\!\!\! &=&\!\!\! \sum_{n=0}^{p-2} c_{p,n} R
f_n(z) - v_p \f{1}{2} \left( m+T -\sqrt{(m+T)^2-4R} \right)
\eqnx
In the above formula the coefficients $c_{p,n}$ are defined in
(\ref{e.defcpn}), while
$v_p$ is just the coefficient of $g_p$ in the constraint $v=0$ (see
eq. (\ref{e.eomv})). Finally the functions $f_n(z)$ are computed
in appendix B and depend on the variable $z(R,T)$, given by (\ref{e.defz}).
In table 1 we present the explicit forms of the
functions $f_n(z)$ for $n \leq 7$.

\begin{table}[t]
\begin{tabular}{ll}
$ f_0(z)=\f{1}{2(z-1)} $ & 
$ f_1(z)=\f{3z-4}{6(z-1)^{3/2}}$ \\
$ f_2(z)=\f{1}{16(z-1)^2}$ &
$ f_3(z)=\f{30z^2-65z+32}{120(z-1)^{5/2}}$ \\
$ f_4(z)=\f{-3z^2+9z-5}{96(z-1)^3}$ &
$ f_5(z)=\f{525z^3-1610z^2+1582z-512}{3360(z-1)^{7/2}}$ \\
$ f_6(z)=\f{-48z^3+168z^2-176z+59}{1536(z-1)^4}$ &
$ f_7(z)=\f{4410z^4-17640z^3+25956z^2-16857z+4096}{40320(z-1)^{9/2}}$
\end{tabular}
\caption{Examples of the functions $f_n(z)$ for small $n$.}
\end{table}

In the final result (\ref{e.ufpfinal}) we see that both the $N_c$ and
$N_f$ dependence is very simple. Moreover the contribution of each
extra flavor enters additively the expression for the moduli. Another
curious feature is the appearance of the original factorization
solutions for the {\em pure} $\nn=2$ $U(N_c)$ gauge theory.
Increasing the number of colors does not change the expressions for
$\upm_p$ in terms of $R$ and $T$.
However the only nontrivial change is encoded in the expression for
$\Lm$ in terms of $R$ and $T$.

If we wanted instead to obtain the effective potential $W(S, \Lambda)$, 
we would have to integrate out $R$ and $T$ from (\ref{fqep}). 
On the field theory side it is very cumbersome how the 
structure of the Seiberg-Witten curve appears in the equations of motion 
for $u_1$ and $\Omega$. On the matrix model, on the other hand, the 
equations of motion for $R$ and $T$ appear naturally to be the same as
ones obtained in the case without flavours. It seems that the particular
combinations of $u_1$ and $\Omega$, embodied in $R$ and $T$, captures
some nontrivial structure of the Seiberg-Witten curves.


\section{Some examples}

In this section we will study some examples to verify that 
the $\uf_p$ we obtained from random matrix models, do factorize 
the appropriate Seiberg-Witten curves with fundamental matter. 

\subsection*{$U(2)$ with 1 flavour}

In this case we have
\eqn
\uf_1\!\!\! &=&\!\!\! 2T- \f{1}{2}\left(m+T-\sqrt{(m+T)^2-4R}\right)\\
\uf_2\!\!\! &=&\!\!\! 2\left(R+\f{T^2}{2}\right) + \f{1}{4}\left(m^2-
2R-T^2+ (T-m) 
\sqrt{(m+T)^2-4R} \right) \\
\Lm^3\!\!\! &=&\!\!\! \f{2R^2}{m+T+\sqrt{(m+T)^2-4R}}
\eqnx
We have verified that with the above choices, the discriminant of the
Seiberg-Witten curve 
\eq
y^2= \left(x^2-u_1 x -u_2 +\f{1}{2} u_1^2 \right)^2 -4\Lm^3 (x+m)
\eqx
vanishes identically, which in the special case of 2 colours proves
factorization. Note that for general $N_c$, complete factorization is
a much stronger condition than the vanishing of the discriminant. 

\subsection*{$SU(N_c)$ with 1 flavour}

In order to consider $SU(N_c)$ theory we have to impose the constraint
that $u_1=0$. Then the parameter $T$ can be expressed in terms of $R$
and the mass $m$ of the additional flavour. Namely we have
\eq
u_1\equiv N_c T -\f{1}{2} \left( m+T -\sqrt{(m+T)^2-4R} \right)=0
\eqx
which gives
\eq
\label{e.tsu}
T=\f{m-\sqrt{m^2-4R +4\f{R}{N_c}}}{2(N_c-1)}
\eqx
Now $R$ is linked directly to the scale of the gauge theory through
\eq
\label{e.lm}
\Lm^{2N_c-1} =\f{R^{N_c}}{\left(m+T+\sqrt{(m+T)^2-4R}\right)/2}
\eqx
where the expression (\ref{e.tsu}) should be used. The remaining
formulas remain however quite complicated functions of $R$, $m$ and
$N_c$. This is in marked contrast to the case of pure gauge theory
without fundamental matter where the passage from $U(N_c)$ to
$SU(N_c)$ is very simple, and $N_c$ enters linearly.  

It is interesting to look at the $m\to \infty$ limit. Then $T\to 0$ as
expected for a pure $\nn=2$ $SU(N_c)$ theory, while (\ref{e.lm})
becomes $\Lm^{2N_c-1}=R^{N_c}/m$. Recall that in pure $SU(N_c)$ theory
$R$ had the interpretation of $\Lm_{pure}^2$. Hence we obtained the
correct field theoretic matching of scales
\eq
\Lm^{2N_c-1} m = \Lm_{pure}^{2N_c}
\eqx
Moreover it is easy to check that then the $\upm_p(R,T,m)$ give a
vanishing contribution as the functions $f_n(z) \to 0$ when $z\to
\infty$ (see appendix B). The above behaviour is quite
clear from the Seiberg-Witten curve perspective. It is reassuring that
it could be also obtained in a simple way from the random matrix
formulas. 

\subsection*{$U(7)$ with 3 flavours}

As a final check of the formulas we considered $U(7)$ theory with 3
flavours with masses $m_1=1$, $m_2=2$, $m_3=3$. For the random choice of
parameters $T=0$ and $R=0.2$ we find $\Lm=0.31643$, while the polynomial
$P_7(x)=x^7+0.45018 x^6-1.3447 x^5-0.5382 x^4+0.5048 x^3+0.16048 x^2
-0.045 x-0.00704$. For the curve
\eq
y^2=P_7(x)^2 -4 \Lm^{11} (x+1)(x+2)(x+3)
\eqx
we find single zeroes at $x=\pm 0.8944$ and a series
of 6 double zeroes in between.

\section{Discussion}

In this paper we used the Dijkgraaf-Vafa proposal linking random
matrix models with fundamental matter and superpotentials for
obtaining explicit formulas for the complete factorization of 
Seiberg-Witten curves for $U(N_c)$ theories with $N_f<N_c$
flavours. These points in the moduli space, forming effectively a
1-parameter manifold, correspond to condensation of all species of
monopoles. As a byproduct we obtained formulas for the solution of the
random matrix model with matter with an arbitrary polynomial
superpotential. 

In order to identify the points in moduli space where the
Seiberg-Witten curve
factorizes we recast the random matrix solution in a way that exhibits      
(i) linearity in the couplings $g_p$ of the deforming tree level
superpotential, (ii) the whole dependence on the glueball superfield
$S$ could be written as a linear coupling of $S$ to a logarithmic
expression. The first property allowed us to identify the moduli space
parameters of the factorized curve $u_p$ as the coefficients of $g_p$,
while the second property is exactly the one found in `integrating-in'
$S$ and thus gave the expression for the gauge theoretic scale $\Lm$
in terms of random matrix model quantities.

The fact that the above procedure works is yet another argument for
the Intriligator-Leigh-Seiberg linearity principle \cite{ILS} and the
validity of `integrating-in'. 
In addition it shows that the matrix model of the Dijkgraaf-Vafa
proposal captures 
quite detailed properties of the field theoretical Seiberg-Witten
curve. In fact we found it surprising that any analytical description
of the very nonlinear complete factorization property could be found
for the case with matter. 

A curious feature of the random matrix
formulas is that the solution for the $u_p$ for the pure $\nn=2$
theory appears {\em linearly} in the complete expression for the theory
with fundamental matter fields. The full `nonlinearity' is encoded in
the formula for $\Lm$ in terms of random matrix parameters. It would
be interesting to understand this structure from the field-theoretical
point of view. 
In addition the random matrix constraints expressed in terms of $R$
and $T$ don't change when adding fundamental matter. They have
precisely the form of equations of motion for the pure $\nn=2$
theory. But now the mapping between $R$ and $T$ and field theoretical
$\Lm$ and $u_1$ becomes complicated. Nevertheless, the question why
the pure $\nn=2$ equations still arise in a disguised form for
theories with fundamental matter poses an interesting question from
the gauge theory perspective.  

There are numerous issues that one could investigate further. The
factorization properties of the pure $\nn=2$ curve are linked with the
concept of master field. This has been investigated in the context of
associated random matrix theory in \cite{GOPAKUMAR}. It would be
interesting to study the factorization formulas obtained in this paper
from a similar point of view, albeit it will surely be much more
involved. 

Another interesting question would be to explore the mathematical
structure linking the random matrix model with matter with
factorization properties of the associated curves. In this paper we
relied heavily on recasting the random matrix expressions guided by
{\em field theoretic} ingredients such as the ILS principle and
integrating-in, for which there is no real direct proof. It would be very
interesting to uncover the {\em mathematical} interrelation between such
seemingly unconnected topics as the factorization of SW curves and
random matrix models.

Finally we hope that the above results could be used for a more
detailed investigation of the physics of  $U(N_c)$ theories with
$N_f<N_c$ flavours along the lines of \cite{DS,FERRARI2}.

\bigskip

\noindent{\bf Acknowledgments} RJ was supported by the EU
network on ``Discrete Random Geometry'' and KBN grant~2P03B09622.
YD was supported by an EC Marie Curie Training site Fellowship at
Nordita, under contract number HPMT-CT-2000-00010.

\medskip

\appendix

\addcontentsline{toc}{section}{Appendices}

\section{Formula for $M(\psi)$}

Since we use explicitly the form of the eigenvalue density in the
variable $\psi$, let us perform the changes of variables
$x=T+2\sqrt{R}\psi$, $w=T+2\sqrt{R}\phi$ in the definition
(\ref{e.defmz}) of $M(x)$:
\eq
\label{e.mpsi}
M(\psi) =\f{1}{2\sqrt{R}} \mbox{\rm Res}_{\phi=\infty}
\f{1}{\phi-\psi} \f{V'(T+2\sqrt{R}\phi)}{\sqrt{\phi^2-1}}
\eqx
Using the power series expansion of the square root
\eq
\f{1}{\sqrt{1-\f{1}{\phi^2}}} =\sum_{k=0}^\infty a_k \f{1}{\phi^{2k}} \equiv
\sum_{k=0}^\infty 2^{-2k} \bin{2k}{k} \f{1}{\phi^{2k}} 
\eqx
it is straightforward to obtain the Laurent expansion of the function
in  (\ref{e.mpsi}):
\eq
\sum_{p\geq 2} g_p \sum_{n=0}^\infty \sum_{k=0}^\infty \psi^n
\f{1}{\phi^{2k+n+2}} a_k \sum_{l=0}^{p-1} \bin{p-1}{l}
\left(2\sqrt{R}\right)^{l-1} T^{p-1-l} \phi^l  
\eqx
{}From this expression we may isolate the coefficient of $1/\phi$ giving
the result quoted in the text:
\eqn
M(\psi) &=& \sum_{p \geq 2} g_p \sum_{n=0}^{p-2} c_{p,n} \psi^n \\
c_{p,n} &=& 2^n R^{\f{n}{2}} \sum_{k=0}^{\left[ \f{p-n-2}{2} \right]}
\bin{2k}{k} \bin{p-1}{2k+n+1} R^k T^{p-n-2-2k} 
\eqnx

\section{Logarithmic integrals $f_n(z)$}

Here we will derive the explicit form of the functions $f_n(z)$
related to the logarithmic integrals
\eq
\label{e.logint}
I_n=\f{2}{\pi} \int_{-1}^1 d\psi \sqrt{1-\psi^2} \cdot \psi^n \cdot
\log \left(1+ 2 \,x\, \psi \right)
\eqx
where $x=\sqrt{R}/(m+T)$. In fact the results simplify significantly
if one reexpresses everything in terms of the variable $z$
\eq
z=\f{m+T}{2R} \left(m+T+\sqrt{(m+T)^2-4R} \right)
\eqx
($x$ is expressed in terms of $z$ as $\sqrt{z-1}/z$). We have to
distinguish two cases:

\subsubsection*{$n$ odd}

The integral (\ref{e.logint}) can be performed using a series
expansion of the logarithm, integrating it term by term using
\eq
\label{e.elemlogint}
\f{2}{\pi} \int_{-1}^1 d\psi \sqrt{1-\psi^2} \cdot \psi^{2n}=
\f{2^{-2n+1}}{(2n+2)} \bin{2n}{n}
\eqx 
and resumming. The result is
\eq
f_n^{odd}(z)=\f{2\sqrt{z-1} \gam{1+\f{n}{2}}}{\sqrt{\pi} z
\gam{\f{5+n}{2}}} \hyp{\f{1}{2}}{1}{1+\f{n}{2}}{\f{3}{2}}{\f{5+n}{2}}{
\f{4(z-1)}{z^2}}
\eqx
For odd $n$, $f_n(z)$ is expressed through elementary functions (see
examples in section 7). It has the form of a polynomial in $z$ of
order $(n+1)/2$ divided by $(z-1)^{\f{n}{2}+1}$.

\subsubsection*{$n$ even}

The integral (\ref{e.logint}) can be again obtained using a
resummation procedure. The result is
\eq
I_n^{even}=-\f{2(z-1) \gam{\f{3+n}{2}}}{\sqrt{\pi} z^2
\gam{3+\f{n}{2}}} \hyp{1}{1}{\f{3+n}{2}}{2}{3+\f{n}{2}}{
\f{4(z-1)}{z^2}}
\eqx

In this case the integral (\ref{e.logint}) involves a logarithm, which
together with the $\log(m+T)$ forms the logarithmic function
multiplying $S$ in (\ref{e.almostgen}). Thus to define the function
$f_n(z)$ for even $n$ we have to subtract from $I_n^{even}$ this
logarithm:
\eq
f_n^{even}(z)\equiv I_n^{even} - \f{2^{-n+1}}{n+2} \bin{n}{\f{n}{2}} \cdot
\log \left(\f{z-1}{z}\right)
\eqx
Its general form turns out to be a polynomial of order $n/2$ divided
by\\ $(z-1)^{\f{n}{2}+1}$.
In table 1 in section 7, for completeness we present the explicit
forms of the functions $f_n(z)$ for $n \leq 7$.


\section{Coefficient of the logarithmic term in (\ref{e.slog})}

In this appendix, we identify the coefficient of the
logarithmic piece of the $\chi=1$ contribution with $S$, as expected from
field theory.
{}From (\ref{e.master}) one can easily read off the coefficient of $\log(m+T)$:
\eq 
4R \int_{-1}^{1} \frac{d\psi}{2 \pi} \sqrt{1 - \psi^2} M(\psi)
\eqx
Using the elementary integral (\ref{e.elemlogint})
one obtains the expression:
\eq
4R \sum_{p \geq 2} g_p \sum_{2l=0}^{p-2} 
\bin{2l}{l} c_{p,2l} \frac{2^{-2l-2}}{l + 1}
\eqx
Inserting the explicit expression for the coefficient $c_{p,2l}$,
leads to:
\eq
\sum_{p \geq 2} g_p \sum_{l=0}^{[\frac{p-2}{2}]}
\sum_{k=0}^{[\frac{p-2l-2}{2}]} \bin{2k}{k} \bin{2l}{l}
\bin{p-1}{2l+2k+1} \frac{1}{l+1} R^{l+k+1} T^{p-2k-2l-2}.
\eqx
Changing to a new summation variable $m = k+l+1$ gives the expression:
\eq
\sum_{p \geq 2} g_p \sum_{m=1}^{[\frac{p}{2}]}
\sum_{l=0}^{m-1}\frac{2m}{p(l+1)} \bin{p}{2m} \bin{2l}{l}
\bin{2m-2l-2}{m-l-1}  R^{m} T^{p-2m}.
\eqx
This is exactly the expression for $S$, provided that:
\eq
2 \sum_{l=0}^{m-1} \frac{1}{l+1} \bin{2l}{l} \bin{2m-2l-2}{m-l-1} = 
\bin{2m}{m}
\eqx
which can be verified.


\begin{thebibliography}{99}

\bibitem{DV1}
R.~Dijkgraaf and C.~Vafa,
``A perturbative window into non-perturbative physics,''
arXiv:hep-th/0208048.

\bibitem{DV2}
R.~Dijkgraaf and C.~Vafa,
``Matrix models, topological strings, and supersymmetric gauge theories,''
Nucl.\ Phys.\ B {\bf 644} (2002) 3
[arXiv:hep-th/0206255].

\bibitem{GEOM}
C.~Vafa,
``Superstrings and topological strings at large N,''
J.\ Math.\ Phys.\  {\bf 42} (2001) 2798
[arXiv:hep-th/0008142].

\bibitem{CIV}
F.~Cachazo, K.~A.~Intriligator and C.~Vafa,
``A large N duality via a geometric transition,''
Nucl.\ Phys.\ B {\bf 603} (2001) 3
[arXiv:hep-th/0103067].

\bibitem{ZANON}
R.~Dijkgraaf, M.~T.~Grisaru, C.~S.~Lam, C.~Vafa and D.~Zanon,
``Perturbative computation of glueball superpotentials,''
arXiv:hep-th/0211017.

\bibitem{DCSW}
F.~Cachazo, M.~R.~Douglas, N.~Seiberg and E.~Witten,
``Chiral rings and anomalies in supersymmetric gauge theory,''
arXiv:hep-th/0211170.

\bibitem{MATTER}
R.~Argurio, V.~L.~Campos, G.~Ferretti and R.~Heise,
``Exact superpotentials for theories with flavors via a matrix integral,''
arXiv:hep-th/0210291.

J.~McGreevy,
``Adding flavor to Dijkgraaf-Vafa,''
arXiv:hep-th/0211009.

H.~Suzuki,
``Perturbative derivation of exact superpotential for meson fields
from  matrix theories with one flavour,''
arXiv:hep-th/0211052.

I.~Bena and R.~Roiban,
``Exact superpotentials in N = 1 theories with flavor and their
matrix  model formulation,''
arXiv:hep-th/0211075.

Y.~Demasure and R.~A.~Janik,
``Effective matter superpotentials from Wishart random matrices,''
arXiv:hep-th/0211082.

B.~Feng,
``Seiberg duality in matrix model,''
arXiv:hep-th/0211202.

B.~Feng and Y.~H.~He,
``Seiberg duality in matrix models. II,''
arXiv:hep-th/0211234.

R.~Argurio, V.~L.~Campos, G.~Ferretti and R.~Heise,
``Baryonic corrections to superpotentials from perturbation theory,''
arXiv:hep-th/0211249.

I.~Bena, R.~Roiban and R.~Tatar,
``Baryons, boundaries and matrix models,''
arXiv:hep-th/0211271.

Y.~Ookouchi,
``N = 1 gauge theory with flavor from fluxes,''
arXiv:hep-th/0211287.

K.~Ohta,
``Exact mesonic vacua from matrix models,''
arXiv:hep-th/0212025.

I.~Bena, S.~de Haro and R.~Roiban,
``Generalized Yukawa couplings and matrix models,''
arXiv:hep-th/0212083.

H.~Suzuki,
``Mean-field approach to the derivation of baryon superpotential from
matrix model,'' 
arXiv:hep-th/0212121.

C.~Hofman,
``Super Yang-Mills with flavors from large N(f) matrix models,''
arXiv:hep-th/0212095.

\bibitem{CV}
F.~Cachazo and C.~Vafa,
``N = 1 and N = 2 geometry from fluxes,''
arXiv:hep-th/0206017.

\bibitem{FERRARI}
F.~Ferrari,
``On exact superpotentials in confining vacua,''
arXiv:hep-th/0210135.

\bibitem{DVP}
R.~Dijkgraaf, S.~Gukov, V.~A.~Kazakov and C.~Vafa,
``Perturbative analysis of gauged matrix models,''
arXiv:hep-th/0210238.

\bibitem{SCHNITZER1}
S.~G.~Naculich, H.~J.~Schnitzer and N.~Wyllard,
``The N = 2 U(N) gauge theory prepotential and periods from a
perturbative matrix model calculation,'' 
arXiv:hep-th/0211123.

\bibitem{SCHNITZER2}
S.~G.~Naculich, H.~J.~Schnitzer and N.~Wyllard,
``Matrix model approach to the N = 2 U(N) gauge theory with matter in
the  fundamental representation,'' 
arXiv:hep-th/0211254.

\bibitem{MOROZOV}
H.~Itoyama and A.~Morozov,
``The Dijkgraaf-Vafa prepotential in the context of general
Seiberg-Witten theory,'' 
arXiv:hep-th/0211245.

\bibitem{FENG}
B.~Feng,
``Geometric dual and matrix theory for SO/Sp gauge theories,''
arXiv:hep-th/0212010.

\bibitem{SW1}
N.~Seiberg and E.~Witten,
``Electric - magnetic duality, monopole condensation, and confinement
in N=2 supersymmetric Yang-Mills theory,'' 
Nucl.\ Phys.\ B {\bf 426} (1994) 19
[Erratum-ibid.\ B {\bf 430} (1994) 485]
[arXiv:hep-th/9407087].

\bibitem{SW2}
N.~Seiberg and E.~Witten,
``Monopoles, duality and chiral symmetry breaking in N=2 supersymmetric QCD,''
Nucl.\ Phys.\ B {\bf 431} (1994) 484
[arXiv:hep-th/9408099].

\bibitem{fa}
P.~C.~Argyres and A.~E.~Faraggi,
``The vacuum structure and spectrum of N=2 supersymmetric SU(n) gauge
theory,'' 
Phys.\ Rev.\ Lett.\  {\bf 74} (1995) 3931
[arXiv:hep-th/9411057].

\bibitem{Klemm}
A.~Klemm, W.~Lerche, S.~Yankielowicz and S.~Theisen,
``Simple singularities and N=2 supersymmetric Yang-Mills theory,''
Phys.\ Lett.\ B {\bf 344} (1995) 169
[arXiv:hep-th/9411048].

\bibitem{DS} 
M.~R.~Douglas and S.~H.~Shenker,
``Dynamics of SU(N) supersymmetric gauge theory,''
Nucl.\ Phys.\ B {\bf 447} (1995) 271
[arXiv:hep-th/9503163].


\bibitem{OBERS}
R.~A.~Janik and N.~A.~Obers,
``SO(N) superpotential, Seiberg-Witten curves and loop equations,''
arXiv:hep-th/0212069.

\bibitem{deBOER}
V.~Balasubramanian, J.~de Boer, B.~Feng, Y.~H.~He, M.~x.~Huang, V.~Jejjala and A.~Naqvi,
``Multi-trace superpotentials vs. Matrix models,''
arXiv:hep-th/0212082.

\bibitem{fgz}
P.~Di Francesco, P.~Ginsparg and J.~Zinn-Justin,
``2-D Gravity and random matrices,''
Phys.\ Rept.\  {\bf 254} (1995) 1
[arXiv:hep-th/9306153].

\bibitem{GrossMigdal}
D.~J.~Gross and A.~A.~Migdal,
``A Nonperturbative Treatment Of Two-Dimensional Quantum Gravity,''
Nucl.\ Phys.\ B {\bf 340} (1990) 333.

\bibitem{ILS}
K.~A.~Intriligator, R.~G.~Leigh and N.~Seiberg,
``Exact superpotentials in four-dimensions,''
Phys.\ Rev.\ D {\bf 50} (1994) 1092
[arXiv:hep-th/9403198].

\bibitem{GOPAKUMAR}
R.~Gopakumar,
``N = 1 theories and a geometric master field,''
arXiv:hep-th/0211100.

\bibitem{FERRARI2}
F.~Ferrari,
``Quantum parameter space and double scaling limits in N = 1 super
Yang-Mills theory,'' 
arXiv:hep-th/0211069.

\end{thebibliography}
\end{document}